# FEED MODEL FOR THE TAPERED SLOT ANTENNAS

[1] Iupikov O.A.

[1] Sebastopol National Technical University, Sebastopol, Ukraine
E-mail: lichne@gmail.com

**Abstract**

In this paper we will describe the feed model for the tapered slot antennas (TSAs) and will find one of the important parameter of this model – transformation coefficient.

## 1. INTRODUCTION

Tapered slot antennas (in particular Vivaldi antennas) are well suited in the applications where a wide frequency band or large scan range are required. To predict the behavior of the real antenna it is reasonable to model one first.

In this paper we will describe the feed model for the tapered slot antennas (TSAs) and will find one of the important parameter of this model – transformation coefficient.

## 2. FEED MODEL

One of the possible feed of the Vivaldi element is shown in Fig.1.

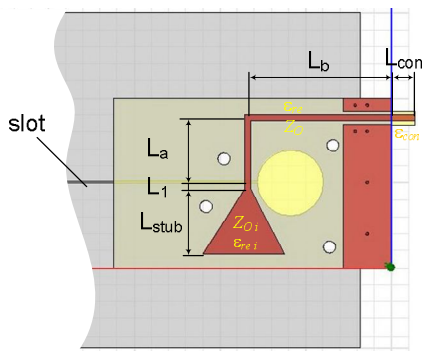

**Fig.1** – Microstrip feed structure

To model the microstrip line to slotline transition we employ an ideal transformer model with transformation coefficient $p$ as described in [1]. The equivalent microwave circuit of this microstrip feed is shown in Fig.2 [4].

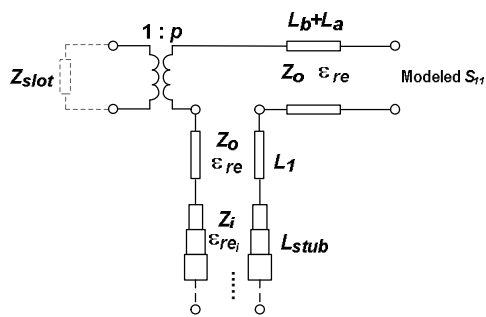

**Fig.2** – The equivalent microwave circuit of microstrip feed

Here $L$ – corresponding to Fig.1 lengths of the microstrip parts;

$Z_o$ – characteristic impedance of the microstrip line;

$Z_{slot}$ – impedance of the slot;

$\varepsilon_{re}$ – the effective dielectric constant of the microstrip line;

$p$ – transformation coefficient.

If $W/h > 2$ then $Z_o$ и $\varepsilon_{re}$ can be calculated by following formulas [1]:

$$Z_o = \frac{377}{\sqrt{\varepsilon_r} \cdot \left[ \frac{W}{h} + 0.883 + \frac{\varepsilon_r + 1}{\pi \varepsilon_r} \cdot \left( \ln\left(\frac{W}{2h} + 0.94\right) + 1.451 \right) + 0.165 \cdot \frac{\varepsilon_r - 1}{\varepsilon_r^2} \right]}$$

$$\varepsilon_{re} = \frac{\varepsilon_r + 1}{2} + \frac{\varepsilon_r - 1}{2} \cdot \frac{1}{\sqrt{1 + 12\frac{h}{W}}}$$

where $\varepsilon_r$ – the permittivity of the dielectric.

In general, these parameters have dependence with frequency. However, if the working frequency is much less then $f_p (\text{GHz}) = 15.66 \frac{Z_0}{h(\text{mils})}$, this dispersion will be neglect [1].

The attenuation in the microstrip line is also exists. This attenuation is caused by two loss component:

conductor loss and dielectric loss. However, in most cases the attenuation coefficient is very small. For example, if we don't take into account losses in our tested model (0.5-2 GHz), this will lead to only 0.2 dB error in simulated S-parameters.

To provide the wideband of the feed the tapered section is used at the end of microstrip line (radial stub). The stub is represented by a number of microstrip lines with equal lengths and different widths.

The characteristic impedances $Z_{o\,i}$ and effective dielectric constants $\varepsilon_{re\,i}$ are different for each section and depend on the corresponding $W/h$ ratio.

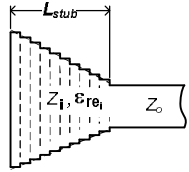

**Fig.3** – Model of the stub

The modeling of the antenna with described feed model can be divided in two steps [2-3]: 1) Electromagnetic modeling of the TSA element or array without feed structure, using for example method of moments; 2) Microwave modeling of the feed structure shown in Fig.2, using four-terminal network theory.

However, to complete second step we have to know transformation coefficient $p$.

## 3. DETERMINATION OF THE TRANSFORMATION COEFFICIENT.

The critical point in the determination of the transformation coefficient is that no accurate analytical expressions exist that can be used in microwave-circuit model. Therefore, an approximate empirical method was employed. This method is based on the curve fitting of the modeled parameter $S_{mk}$ of the microstrip structure with one unknown $p$ to the measurements. This can be done iteratively for many values of $p$ until the minimum difference between the modeled and measured parameters will be achieved. For the measurements the one element or small array can be produced.

For the case of the curve fitting to the phase, additional unknown parameter, the length of the connector $L_{con}$ was introduced. The reason for this is that during measurements, the position of the reference plane is not accurately known, but can have a significant effect on the measured phase.

Thus, the algorithm of finding $p$ is following:

- perform the modeling of the structure without feed in Electromagnetic (EM) simulator (only once);
- then, the Microwave simulator is used to calculate $S_{mk}$ of the considered structure for the specified range of $p$ (and for each value of $L_{con}$ for the case of the fitting to the phase). This step is performed for all frequency points. As the result, we get a multi-dimensional matrix of S-parameters for the range of $p$, $L_{con}$ and frequencies.
- calculate the correlation matrix (for the magnitudes of S-parameters) and error matrix (for phases of S-parameters) with formulas:

$$R(p,L_{con}) = \frac{\text{cov}\left(\left|S_{mk}^{Model}(p,L_{con},f)\right|,\left|S_{mk}^{Meas}(f)\right|\right)}{\sqrt{\text{cov}\left(\left|S_{mk}^{Model}(p,L_{con},f)\right|,\left|S_{mk}^{Model}(p,L_{con},f)\right|\right)\cdot\text{cov}\left(\left|S_{mk}^{Meas}(f)\right|,\left|S_{mk}^{Meas}(f)\right|\right)}}$$

$$Err_{PH}(p,L_{con}) = \sum_{f=F\min}^{F\max}\left|\text{angle}\left(S_{mk}^{Model}(p,L_{con},f)\right)-\text{angle}\left(S_{mk}^{Meas}(f)\right)\right|$$

where $\text{cov}\left(\left|S_{mk}^{Model}\right|,\left|S_{mk}^{Meas}\right|\right)$ – the covariance coefficient between $S_{mk}^{Model}(p,L_{con},f)$ and $S_{mk}^{Meas}(f)$;

- Afterwards, the minimal error $\min\{Err(p,L_{con})\}$ and the maximum correlation $\max\{R(p,L_{con})\}$ are found for each set of parameters $p$ and $L_{con}$. This set corresponds to the optimal transformation coefficient and connector length. It should be noted that for the case of the fitting to the magnitude, the optimal value $p_{opt}$ is the same for the entire range of $L_{con}$, as it is independent on this parameter.

The criterion of the maximum cross-correlation coefficient $R(p,L_{con})$ seems to work better than the minimum difference error $\min\{Err(p,L_{con})\}$ for the case of the fitting to the magnitude. The reasoning for this is the nature of the correlation coefficient which "compares" the shapes of the curves and the positions of the maximums and minimums; and since the magnitude of $S_{mk}$ (especially reflection coefficients $S_{mm}$ of array elements) typically has a pronounced resonance form, this criterion leads to a better agreement between the modeled and measured parameters in terms of the shape (the number and positions of the resonances).

However, the use of the maximum cross-correlation coefficient as the criterion does not lead to a good result when the fitting is performed for the phase. This can be explained by a typically liner-varying behavior of the phase over the frequency band, so the cross-correlation coefficient changes slowly when both the modeled and measured phase functions are linear (even if the second parameter $L_{con}$ can differ a lot).

Figure Fig.4 show results of the optimization for two unknown parameters ($p$ and $L_{con}$) for the 3-element Vivaldi array by using reflection coefficient $S_{11}$.

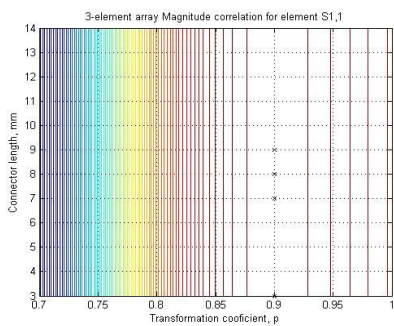

**Fig.4a** – Correlation coefficient between measured and simulated magnitudes of $S_{11}$

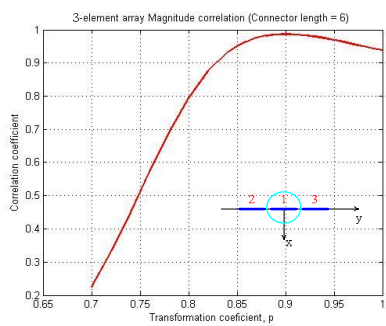

**Fig.4b** – Cross-section of Fig.4a with $L_{con}$ = 6mm

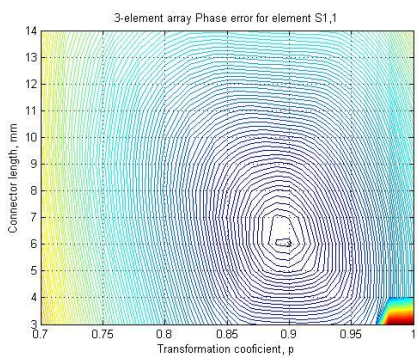

**Fig.4c** – Average phase error between measured and simulated phase of $S_{11}$

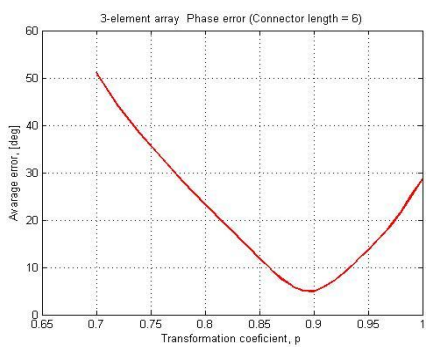

**Fig.4d** – Cross-section of Fig.4c with $L_{con}$ = 6mm

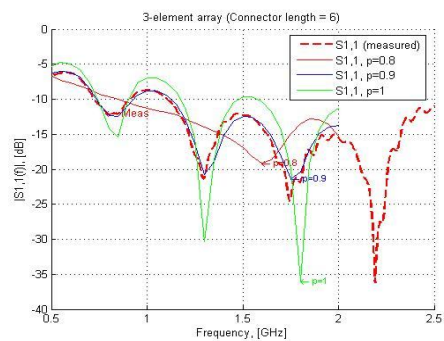

**Fig.4e** – Measured (dotted) and simulated for different $p$ magnitude of $S_{11}$

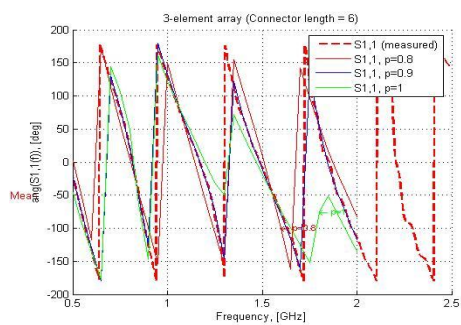

**Fig.4f** – Measured (dotted) and simulated for different $p$ phase of $S_{11}$

Thus, it was found that transformation coefficient for this array structure is equal $p = 0,9$ for both magnitude and phase curve fitting. And we obtained good agreement between measurements and modeling with $p = 0,9$.

### REFERENCES

1. K.C. Gupta, et. al., Microstrip lines and slotlines, Artech House, Norwood, MA, 1979.
2. D.H. Schaubert, Endfire slotline antennas, JINA conf., Journees internationals de Nice sur les Antennes, Nice, France, Nov., 1990, p.253-265.
3. A.B. Smolders, M.J. Arts, Wide-band antenna element with integrated balun, IEEE Int Symp., Atlanta, USA, 1998.
4. M. V. Ivashina, E. A. Redkina, and R. Maaskant, 'An Accurate Model of a Wide-Band Microstrip Feed for Slot Antenna Arrays', The 2007 IEEE International Symposium on Antennas and Propagation, Honolulu, Hawaii, USA June 10-15, 2007.